\documentclass[12pt]{article}  
\usepackage{amsfonts}   

\usepackage{achicago} 

\begin{document}

\date{New Developments in Logic and Philosophy of Science.  Proceedings of SILFS 2014: Societ\`{a} Italiana di Logica e Filosofia delle Scienze\\University of Rome `Roma TRE', 18-20 June 2014} 
\sloppy

\title{\vspace{-.5in} Historical and Philosophical Insights about General Relativity and Space-time from Particle Physics\footnote{Published version is in \emph{New Directions in Logic and the Philosophy of Science}.  Proceedings of SILFS 2014: Società Italiana di Logica e Filosofia delle Scienze, University of Rome 'Roma TRE', 18-20 June 2014, edited by Laura Felline, Antonio Ledda, Francesco Paoli, and Emanuele Rossanese.  College Publications, London (2016), pp. 291-301.  http://www.collegepublications.co.uk/downloads/silfs00029.pdf.  One reference added. } \\} 

\author{J. Brian Pitts  \\  University of Cambridge, jbp25@cam.ac.uk \\   John Templeton Foundation grant \#38761 } %

\maketitle



\abstract{Historians recently rehabilitated Einstein's ``physical strategy'' for General Relativity (GR).  Independently, particle physicists similarly re-derived Einstein's equations for a massless spin 2 field. But why not a light \emph{massive} spin 2, like Neumann and Seeliger did to Newton?  Massive gravities are bimetric, supporting  conventionalism over  geometric empiricism.  Nonuniqueness lets field equations explain geometry but not \emph{vice versa}.  Massive gravity would have blocked Schlick's critique of Kant's synthetic \emph{a priori}. Finally in 1970 massive spin 2 gravity seemed unstable or empirically falsified. GR was vindicated, but later and on better grounds. 
However, recently dark energy and theoretical progress have made massive spin 2 gravity potentially viable again.}



\pagebreak 



\section{Einstein's Physical Strategy Re-Appreciated by GR Historians}

 Einstein's General Relativity is often thought to owe much to his  various principles (equivalence, generalized relativity, general covariance, and Mach's) in contexts of discovery and justification. But a prominent result of the study of Einstein's process of discovery is a new awareness of and appreciation for Einstein's \emph{physical strategy}, which coexisted with his mathematical strategy involving various thought experiments and principles.  The  physical strategy had as some key ingredients  the  Newtonian limit, the electromagnetic analogy, coupling of all energy-momentum \emph{including gravity's} as a source for gravity, and energy-momentum conservation as a consequence of the gravitational field equations \emph{alone} \cite{Janssen,BradingConserve,Renn,RennSauer,JanssenRenn,RennSauerPathways}.  
 Einstein's mathematical strategy sometimes is seen to be less than compelling \cite{NortonEliminative,StachelEliminative}, leaving space that one might hope to see filled by the physical strategy.

It has even been argued recently, contrary to longstanding views rooted in Einstein's post-discovery claims  \cite{Feynman}, that he found his field equations using his  physical strategy \cite{JanssenRenn}.   Just how the physical strategy led to the field equations is still somewhat mysterious, resisting rational reconstruction \cite{RennSauerPathways}.


\section{Particle Physicists  Effectively Reinvent Physical Strategy}

There is, however, an enormous body of relevant but neglected physics literature from the 1920s onward.  In the late 1930s progress in particle physics led to Wigner's  taxonomy of relativistic wave equations in terms of mass and spin.
``Spin'' is closely related to tensor rank; hence spin-0 is a scalar field, spin-1 a vector, spin-2 a symmetric tensor.
``Mass'' pertains to the associated ``particles'' (quanta) of the field (assuming that one plans to quantize).  (The constants $c$ and  $\hbar$ are set to $1$.) 
Particle masses are related inversely to the range of the relevant potential, which for a point source takes the form $\frac{1}{r} e^{-mr}$.  
 Hence the \emph{purely classical concepts} involved are merely wave equations (typically second order) that in some cases also have a new fundamental inverse length scale permitting algebraic, not just differentiated, appearance of the potential(s) in the wave equation---basically the Klein-Gordon equation.  Despite the facade of quantum terminology---there is no brief equivalent of ``massive graviton''---much of particle physics literature is the \emph{systematic exploration of classical field equations covariant under (at least) the Poincar\'{e} group} distinctive of Special Relativity---though the larger 15-parameter conformal group or the far more general `group' of transformations in General Relativity are not excluded.  Hence drawing upon particle physics literature is simply what eliminative induction requires for classical field theories.   
 
In this context, Fierz and Pauli  found in 1939 that the linearized vacuum Einstein equations are just the equations of a massless spin-2 field \cite{FierzPauli}. Could Einstein's equations be derived from viewpoints in that neighborhood?  Yes:   arguments were devised to the effect that, assuming \emph{special} relativity and some standard criteria for viable field theories (especially stability), along with the empirical fact of light bending, Einstein's equations were the unique result---what philosophers call an  
eliminative induction  \cite{Kraichnan,Gupta,Feynman,Weinberg64c,OP,Deser,VanN,BoulangerEsole}.  
The main freedom lay in including or excluding a graviton mass.

If particle physicists effectively reinvented Einstein's physical strategy, how did they get a unique result, in contrast to the residual puzzles found by Renn and Sauer \cite{RennSauerPathways}?  The biggest difference is a new key ingredient, the elimination of negative energy degrees of freedom, which threaten stability.  Eliminating negative energy degrees of freedom nearly fixes the linear part of the theory  \cite{VanN}, and fixes it in such a way that the nonlinear part is also fixed almost uniquely.  Technical progress in defining energy-momentum tensors also helped.  Such derivations bear a close resemblance to Noether's converse Hilbertian assertion \cite{Noether}---an unrecognized similarity that might have made particle physicists' job easier.


\section{How Particle Physics Could Have Helped Historians of GR}

The main difficulty in seeing the similarity between Einstein's physical strategy and particle physicists' spin-2 derivation of Einstein's equations is the entrenched habits of mutual neglect between communities. 
If one manages to encounter both literatures, the resemblance is evident.  Particle physics derivations subsume Einstein's physical strategy especially as it appears in the little-regarded \emph{Entwurf}, bringing it to successful completion with the correct field equations, using weaker and hence more compelling premises.  Thus the \emph{Entwurf} strategy really was viable in principle.  In particular, Einstein's appeal to the principle of energy-momentum conservation  \cite{EinsteinEntwurf,NortonField,BradingConserve} contains the key ingredient that makes certain particle physics-style derivations of his equations successful  \cite{SliBimGRG},
 namely, that the gravitational field equations alone should entail conservation, without use of the material field equations.  Later works  derived that key ingredient as a \emph{lemma} from gauge invariance, arguably following from positive energy, arguably following from stability.   Einstein's equations follow rigorously  from special relativistic classical field theory as the simplest possible local theory of a massless field that bends light and that looks stable by having positive energy  \cite{VanN} (or maybe one can admit only a few closely related rivals); van Nieuwenhuizen overstated the point only slightly in saying that ``general relativity follows from special relativity by excluding ghosts'' (negative-energy degrees of freedom) \cite{VanN}. 
Excluding ghosts nearly fixes the linear approximation. If one does not couple the field to any source, it is physically irrelevant.  If a source is introduced, the linearized Bianchi identities lead to inconsistencies unless the source is conserved.  The only reasonable candidate is the total stress-energy-momentum, including that of gravity. As a result the initial flat background geometry merges with the gravitational potential, giving an effectively geometric theory, hence with Einstein's nonlinearities \cite{Kraichnan,Deser,SliBimGRG}. 
More recently Boulanger and Esole commented that 
\begin{quote} it is well appreciated that general relativity is the unique way to consistently deform the Pauli-Fierz action $\int\mathcal{L}_2$ for a free massless spin-2 field under the assumption of locality, Poincar\'{e} invariance, preservation of the number of gauge symmetries and the number of derivatives \cite{BoulangerEsole}.  \end{quote}

Familiarity with the particle physics tradition would have shown historians of GR that Einstein's physical strategy was in the  vicinity of a compelling argument for his `correct' field equations.  
  Hence it would not be  surprising if his physical strategy played an important role in Einstein's process of discovery and/or justification.   Might historians of GR not thus have re-appreciated Einstein's physical strategy decades earlier?  Might the apparent tortuous reasoning \cite{RennSauerPathways}  regarding just how Einstein's physical strategy leads to Einstein's equations have been brought into sharper focus, with valid derivations available to compare with Einstein's trail-blazing efforts?  Let $POT$ be the gravitational potential, $GRAV$ a second-order differential operator akin to the Laplacian, and $MASS$ be the total stress-energy-momentum, which generalizes the Newtonian mass density  \cite{Renn}.  Whereas the schematic equation $GRAV(POT)=MASS$ is supposedly innocuous, particle physics would also expose the gratuitous exclusion of a mass term, which would require the form $GRAV(POT) +POT=MASS.$


\section{Massive Gravities?}


 One might expect that a light massive field of spin-$s$ would approximate a massless spin-$s$ field as closely as desired, by making the mass small enough.  Hugo von Seeliger in the 1890s already clearly made a similar point; he wrote  (as translated by Norton) that Newton's law was ``a purely empirical formula and assuming its exactness would be a new hypothesis supported by nothing.'' \cite{Seeliger1895a,NortonWoes}  With the intervention of Neumann, which Seeliger accepted, the exponentially decaying point mass potential later seen as characteristic of massive fields was also available in the 1890s.  (No clear physical meaning was available yet, however).  It is now known that this expectation of a smooth massless limit is true for Newtonian gravity,  relativistic spin-0 (Klein-Gordon), spin-$1/2$ (Dirac), a single spin-1 (de Broglie-Proca massive electromagnetism, classical and quantized), and, in part, a Yang-Mills spin-1 multiplet (classically, but not when  quantized) \cite{DeserMass}.  Hence the idea that gravity might have a finite range due to a non-zero `graviton mass' was not difficult to conceive.  Indeed Einstein reinvented much of the idea in the opening of his 1917 cosmological constant paper \cite{EinsteinCosmological}, intending it as an analog of his cosmological constant.  Unfortunately Einstein erred, forgetting the leading zeroth order term  \cite{Heckmann,FMS,NortonWoes,HarveySchucking}. Plausibly, Einstein's mistaken analogy helped to delay conception of doing to GR what Seeliger and Neumann had done to Newton's theory.

Particle physicists would not be much affected by Einstein's  mistake, however; Louis de Broglie  entertained massive photons from  1922 \cite{deBroglieBlack}, and the Klein-Gordon equation would soon put the massive scalar field  permanently on the map as a toy field theory.  Particle physicists got an occasion to think about gravity when a connection  between Einstein's theory and the rapidly developing work on relativistic wave equations appeared in the late 1930s \cite{FierzPauli}.  From that time massive gravitons saw sustained, if perhaps not intense, attention until  1970   \cite{TonnelatSecond,PetiauCR41a,deBroglie1,DrozVincentMass59,OP,FMS}.

One would expect that anything that can be done with a spin-2, can be done more easily with spin-0.  Thus the Einstein-Fokker geometric formulation of Nordstr\"{o}m's theory (massless spin-0) is a simpler (conformally flat) exercise in Riemannian geometry than  Einstein's own theory.  There are also many massive scalar gravities \cite{PittsScalar},  and by analogy \cite{OP}. 
 The scalar case, though obsolete, is interesting not only because it is easy to understand, but also because massive scalar gravities manifestly make sense as  classical field theories.  While massive scalar gravity has not been an epistemic possibility since 1919 (the bending of light), it ever remains a metaphysical possibility.  Thus  the modal lessons about multiple geometries are not hostage to the changing fortunes of massive spin-2 gravity.
Massive scalar gravity also shows that (\emph{pace} \cite[p. 179]{MTW}  \cite{NortonNordstrom}) gravity did not have to burst the bounds of special relativity on account of Nordstr\"{o}m's theory having the larger 15-parameter conformal group; massive scalar gravities have just the 10-parameter Poincar\'{e} group of symmetries.


\section{Explanatory Priority of Field Equations over Geometry}

In GR, the power of Riemannian geometry to determine the field equations tempts one to think that geometry generically is a good explanation of the field equations.  Comparing GR with its massive cousins sheds crucial light on that expectation.

A key fact about massive gravities is the non-uniqueness of the mass term \cite{OP}, in stark contrast to the uniqueness of the kinetic term (the part that has derivatives of the gravitational potentials), which matches Einstein's theory.  The obvious symmetry group for most massive spin-2 gravities is just the Poincar\'{e} group of special relativity \cite{OP,FMS}; the graviton mass term breaks general covariance.  If one wishes nonetheless to recover formal general covariance, then a graviton mass term must introduce a background metric \emph{tensor} (as opposed to the numerical matrix $diag(-1,1,1,1)$ or the like), typically (or most simply)  flat.

The ability to construct many different field equations from the same geometrical ingredients supports the dynamical or constructive view of space-time theories \cite{BrownPhysicalRelativity,ButterfieldCausalityConventionGeometry}.  The opposing space-time realist view holds that the geometry of space-time instead does the explaining.  
According to the realist conception of Minkowski  spacetime, 
\begin{quote}
(2)  The spatiotemporal interval $s$ between events $(x, y, z, t)$ and $(X, Y, Z, T)$  along  a straight [footnote suppressed]  line connecting  them  is a property  of the spacetime, independent  of the matter it contains, and is given by 
 $$ s^2 = (t - T)^2  - (x - X)^2  - (y - Y)^2 - (z - Z)^2. \hspace{.5in}	(1) $$
 When $s^2  > 0,$ the interval $s$ corresponds  to times elapsed on an ideal clock; when $s^2  < 0$, the interval $ s$ corresponds  to spatial  distances measured by ideal rods (both employed in the standard way). \cite{NortonFails} 
\end{quote}
One might worry that the singular noun ``[t]he spatiotemporal interval'' is worrisomely  ambiguous, as is the adjective ``straight.'' Why can there be only one metric?  
Resuming:  
\begin{quote}(3)  Material  clocks and rods measure these times and distances because the laws of the matter  theories that  govern them are adapted  to the independent  geometry of this spacetime.  \cite{NortonFails} \end{quote}
 But (3) is \emph{false} for massive scalar gravity, in which matter $u$ sees $g_{\mu\nu},$ not the flat metric $\eta_{\mu\nu}$, as is evident by inspection of the matter action $S_{matter}[g_{\mu\nu}, u]$ \cite{Kraichnan},
 which lacks $\sqrt{-\eta}, $ the volume element of the flat metric.  
Unlike space-time realism, constructivism makes room for Poincar\'{e}-invariant field theories in which rods and clocks do not see the flat geometry, such as massive scalar gravities. 

 Even if one decides somehow that massive scalar gravities, despite being just Poincar\'{e}-invariant, are not theories in Minkowski space-time, thus averting the falsification of space-time realism, it still fails on modal grounds.   It simply takes for granted that the world is simpler than we have any right to expect, neglecting a vast array of metaphysical possibilities, some of them physically interesting.  Space-time realism, in short, is modally provincial.  Norton himself elsewhere decried such narrowness in a different context:  one does not want a philosophy of geometry to provide a spurious apparent necessity to a merely contingent conclusion that GR is the best space-time theory \cite[pp. 848, 849]{Norton}.   Constructivism, like conventionalism  \cite[pp. 88, 89]{PoincareFoundations} \cite{BenMenahemPoincare,GrunbaumEarman,WeinsteinScalar}, does not assume that there exists a unique geometry; space-time realism, like the late geometric empiricism of Schlick and Eddington, does assume a unique geometry.   It is striking  that critiques of conventionalism also have usually ignored the possibility of multiple geometries \cite{PutnamConventionLong,SpirtesConvention,FriedmanFoundations,Torretti,ColemanKorteHarmonic,NortonConvention}.


\section{Massive Gravity as Unconceived Alternative}

  The problem of unconceived alternatives or underconsideration \cite{SklarUnborn,vanFraassenLS,StanfordUnconceived}  can be a serious objection to scientific realism. Massive scalar gravity posed such a problem during the 1910s.  Massive spin-2 gravities  continued to pose such a problem for philosophers and general relativists at least until 1972, when the unnoticed threat went away. \emph{C.} 1972 a dilemma appeared:  massive spin-2 gravity was either empirically falsified in the pure spin-2 case because of a discontinuous limit of small \emph{vs.} $0$ graviton mass (van Dam-Veltman-Zakharov discontinuity), or it was violently unstable for the spin 2-spin 0 case because the spin-0 has negative energy, permitting spontaneous production of  spin-2 and spin-0 gravitons out of nothing.   Particle physics gives, but it can also take away.
More recently particle physics has given back, reviving the threat to realism about GR due to unconceived alternatives.
 While underdetermination by approximate but arbitrarily close empirical equivalence has long been clear in electromagnetism \cite{UnderdeterminationPhoton}, 
it is now  (back) in business for gravitation as well.

For philosophers and physicists interested in space-time prior to 1972, or since 2010, not conceiving of massive gravity means suffering from failure to entertain a rival to GR that is \emph{a priori} plausible (a decently high prior probability $P(T)$ if one is not biased against such theories, and if the smallness of the graviton mass does not seem problematic), has good fit to data (likelihoods 
$P(E|T)$ approximating those of GR), and, crucially, has significantly different philosophical consequences from GR.

The underdetermination suggested by massive gravities and massive electromagnetism is weaker in four  ways than the general  thesis  often discussed:   it is  restricted to mathematized sciences, is  defeasible rather than algorithmic in generating the rivals, involves  a  one-parameter family of rivals that work as a team rather than a single rival theory, and is  asymmetric:   the family (typically) remains viable as long as the massless theory is, but not \emph{vice versa}.  


\section{Schlick's Critique of Kant's Synthetic \emph{A Priori} }

The years around 1920 were crucial for a rejection of even a broadly Kantian \emph{a priori} philosophy of geometry, especially due to Moritz Schlick's influence \cite{Schlick,SchlickCritical,Coffa,BitbolConstituting,DicksonKantFriedman}, and saw a partial retreat from conventionalism toward geometric empiricism \cite{HowardEinsteinSchlick,Ryckman,WalterSchlickPoincare}.
Schlick argued  that GR made even a broadly Kantian philosophy of geometry impossible because the physical truth about the actual world was incompatible with it \cite{Schlick,SchlickCritical,Ryckman,Coffa}. Coffa agreed, stuffing half a dozen success terms into two paragraphs in praise of Schlick \cite[pp. 196, 197]{Coffa}. 
That Schlick, brought up as a physicist under Planck, could, in principle, have done to Nordstr\"{o}m's  and Einstein's theories what Neumann, Seeliger and Einstein had done to Newton's, thus making room for synthetic \emph{a priori} geometry, seems not to have been entertained.  Neither was the significance of the 1939 work of Fierz and Pauli \cite{FierzPauli}.  

Recognizing massive gravities as unconceived alternatives, one views Schlick's work in a different light. 
 Schlick argued  that General Relativity either falsifies or evacuates Kant's synthetic \emph{ a priori} \cite{SchlickCritical}. He then quit thinking about space-time, and was assassinated in 1936.
 But post-1939, the flat background geometry \emph{present in the field equations of massive gravity} would leave a role for Kant's geometrical views even in modern physics after all. (This multi-metric possibility is \emph{not} the old L\"{ot}ze move of retaining flat geometry \emph{via} universal forces! Such entities cannot be independently identified, and turn out to be even more arbitrary than one might have expected due to a new gauge freedom \cite{Grishchuk,NortonConvention}.  The observability of the flat metric, indirect though it is, makes the difference \cite{FMS}.  One can ascertain the difference between the two geometries, which is the gravitational potential.) More serious trouble for Kant would arise finally when the van Dam-Veltman-Zakharov discontinuity was discovered. Hence Kant was viable until 1972, not 1920!---and maybe again today.

Massive gravities also bear upon Friedman's claim that the equivalence principle (viewed as identifying gravity and inertia) in GR is constitutively \emph{a priori}, that is, required for this or similar theories to have empirical content  \cite{FriedmanDynamicsReason}.  Massive gravities, if the limit of zero graviton mass is smooth as least (true for spin-0, recently arguable for spin-2), have empirical content that closely approximates Nordstr\"{o}m's and Einstein's theories, respectively, while the massive spin-0 and (maybe) massive spin-2 sharply distinguish gravity from inertia.  The empirical content resides not in  principles or in views about geometry, but in partial differential field equations \cite{FMS,BrownPhysicalRelativity}.


\section{Recent Breakthrough in Massive Gravity}

In the wake of the seemingly fatal dilemma of 1972, massive gravity was largely dormant  until the late 1990s.  Then it started to reappear   due to the ``dark energy''   phenomenon indicating that the  cosmic expansion is accelerating, casting doubt on the long-distance behavior of GR---the regime where a graviton mass term should be most evident.  A viable massive gravity theory must, somehow, achieve a smooth massless limit in order to approximate GR, and be stable (or at least not catastrophically unstable). 
That such an outcome is possible is now often  entertained. 
Massive gravity is now a ``small industry'' \cite[p. 673]{HinterbichlerRMP}  and is worthy of notice by philosophers of science.

Since  2000, Vainshtein's early argument that the van Dam-Veltman-Zakharov discontinuity was an artifact of an approximate rather than exact solution procedure was revived and generalized \cite{Vainshtein,Vainshtein2,DeffayetRecoveryPRD}.  Thus pure spin-2 gravity might have a continuous massless limit after all, avoiding empirical falsification.  The other problem was that an exact rather than merely approximate treatment of massive gravity shows, apparently, all versions of pure spin-2 gravity at the lowest level of approximation, are actually spin 2-spin 0 theories, hence violently unstable, when treated exactly \cite{DeserMass}.  This problem was solved by a theoretical breakthrough in late 2010, where it was found how to choose nonlinearities and carefully redefine the fields such that very special pure spin-2 mass terms at the lowest (linear) approximation \emph{remain pure spin-2 when treated exactly} \cite{deRhamGabadadze,HassanRosen}.

The answers to deep questions of theory choice and conceptual lessons about space-time theory depend on surprises found in sorting out fine technical details in current physics literature.  Thus philosophers should not  assume that all the relevant physics has already been worked out long ago and diffused in textbooks.  Lately things have changed rather rapidly, with  threats of  reversals \cite{DeserWaldronAcausality}. 
Getting the smooth massless limit \emph{via} the Vainshtein mechanism is admittedly ``a delicate matter'' (as a referee nicely phrased it) \cite{deRhamLRR}.

One  needs to reexamine all the conceptual innovations of GR that, by analogy to massive electromagnetism, one would expect to fail in massive gravity \cite{FMS}.  Unless they reappear in massive gravity, or massive gravity fails again, then such innovations are optional. Surprisingly many  of those innovations \emph{do reappear} if one seeks a consistent notion of causality \cite{MassiveGravity1}, including gauge freedom, making those the robust and secure conceptual innovations---whether or not massive gravity survives all the intricate questions that have arisen recently.  If massive gravity fails, then General Relativity's conceptual innovations are required.  If massive gravity remains viable, then General Relativity's conceptual innovations are required only insofar as they also appear in massive gravity.   It is striking how the apparent philosophical implications can change with closer and closer investigation.  %

\end{document}